\documentclass{eas}
\usepackage{graphicx}
\newcommand{\meanAV}{\langle A_{\rm V} \rangle}
\newcommand{\meanW}{\langle W_{\rm_{CO}} \rangle}

\begin{document}

\title{Understanding the physics of the X-factor} 
\runningtitle{Understanding the physics of the X-factor}
\author{S.~C.~O.~Glover}\address{Zentrum f\"ur Astronomie der Universit\"at Heidelberg, Institut f\"ur Theoretische
Astrophysik, Albert-Ueberle-Str.\ 2, 69120 Heidelberg, Germany}
\author{M.-M. {Mac Low}}\address{Department of Astrophysics, American Museum of Natural History, Central Park West at 79th Street, New York, NY 10024, USA}
\begin{abstract}
We study the relationship between the H$_2$ and CO abundances in simulated molecular clouds
using a fully dynamical model of magnetized turbulence coupled to a detailed chemical network. 
We find that the CO-to-H$_{2}$ conversion factor for a given molecular cloud, the so-called X-factor,
is determined primarily by the mean extinction of the cloud, rather than by its metallicity. Our results explain the discrepancy observed in low metallicity systems between cloud masses derived from CO observations and other techniques such as infrared emission, and predict that CO-bright clouds in 
low metallicity systems should be systematically larger and/or denser than Milky Way clouds.
\end{abstract}
\maketitle
\section{Introduction}

Observed star formation takes place within giant molecular clouds  
(GMCs), so understanding how these clouds form and evolve is a key step towards
understanding star formation. The main chemical constituent of
any GMC is molecular hydrogen (H$_{2}$), but this is very difficult to observe {\em in situ}.  
For this reason it is common to use emission from carbon monoxide (CO) as a proxy for
H$_{2}$. 

In order to do this, however, it is necessary to understand the relationship between the 
distributions of the H$_{2}$ and the CO. They have rather different formation mechanisms:
H$_{2}$ forms on the surface of dust grains,
while CO forms in the gas-phase as a product of  ion-neutral chemistry. Both are 
readily photodissociated by ultraviolet (UV) radiation, but H$_{2}$ can protect itself
from this radiation via self-shielding even in relatively low column density gas 
(Draine \& Bertoldi \cite{db96}). We therefore expect to find large variations in the CO/H$_{2}$ 
ratio within any given GMC, with the H$_{2}$ filling a significantly larger volume of the cloud than 
the CO.

Despite this, there is good evidence that CO {\em emission} is a good tracer of
H$_{2}$ mass within the Milky Way (see e.g.\ Solomon {\em et~al.\/} \cite{sol87}, 
Dame {\em et~al.\/} \cite{dame01}).
A number of independent studies have shown that GMCs in the Galactic disk show
a good correlation between  the integrated intensity of the $J = 1 \rightarrow 0$ rotational 
transition line of $^{12}$CO and the H$_{2}$ column density. This correlation is typically
described in terms of a conversion factor $X_{\rm CO}$ (the `X-factor'), given by
\begin{equation}
X_{\rm CO} = \frac{N_{\rm H_{2}}}{W_{\rm CO}} \simeq 2 \times 10^{20} {\rm cm^{-2} \: K^{-1} \: km^{-1} \: s},
\end{equation}
where $W_{\rm CO}$ is the velocity-integrated intensity of the CO $J = 1 \rightarrow 0$ 
emission line, averaged over the projected area of the GMC, and $N_{\rm H_{2}}$ is the 
mean H$_{2}$ column density of the GMC, averaged over the same area.

However, the issue of the environmental dependence of $X_{\rm CO}$ remains highly
contentious. Extragalactic measurements of $X_{\rm CO}$ that use a virial analysis to determine 
cloud masses find values for $X_{\rm CO}$ that are similar to those obtained in the Milky Way, with 
at most a weak metallicity dependence (e.g.\ Rosolowsky {\em et~al.\/} \cite{ros03}, Bolatto {\em 
et~al.\/} \cite{bolatto08}). On the other hand, measurements that constrain GMC masses using techniques that do not depend on CO emission consistently find values for  $X_{\rm CO}$ that 
are much larger than the Galactic value and that are suggestive of a strong metallicity dependence 
(e.g.\ Israel \cite{israel97},  Leroy {\em et~al.\/} \cite{leroy09}).

Numerical simulations provide us with one way to address this observational dichotomy. 
If we can understand the distribution of CO and H$_{2}$ in realistic models of GMCs, then we 
may begin to understand why the different types of observation give such different results.
In this contribution, we summarize the results from some of our recent numerical simulations 
that self-consistently model both the chemistry  and the turbulent dynamics of the gas within 
GMCs, and discuss what they can tell us about the behaviour of $X_{\rm CO}$.

\section{Method}
We have performed a large number of simulations of the chemical and thermal 
evolution of the turbulent, dense interstellar medium using a modified version of
the ZEUS-MP magnetohydrodynamical code. Our modifications include the
addition of a simplified treatment of hydrogen, carbon and oxygen chemistry,
a detailed atomic and molecular cooling function, and a treatment of the effects 
of ultraviolet radiation using a six-ray approximation. Full details of these modifications
can be found in Glover {\em et al.\/}~(\cite{glo10}).

Our simulations begin with initially uniform atomic gas, threaded by a uniform magnetic field 
with strength $B_{0} = 5.85 \mbox{ }\mu {\rm G}$. The initial velocity field is turbulent, with
power concentrated on large scales, and with an initial rms velocity of 5~${\rm km} \: {\rm s^{-1}}$. 
We drive the turbulence so as to maintain approximately the same rms velocity throughout the simulation. We adopt periodic boundary conditions for the gas and in most cases use a cubical
simulation volume with a side length $L= 20$~pc. In a few simulations, we adopt a smaller-sized
box, with $L = 5$~pc. We have run simulations with a variety of mean densities and metallicities,
in order to span a range of different physical conditions. Full details of these simulations can be
found in Glover \& {Mac Low} (\cite{gm10}). 

\section{Results}
Because the CO in many of our simulations is optically thick, an accurate determination
of $W_{\rm CO}$ would require a full non-LTE radiative transfer calculation, a complex undertaking
outside the scope of our present study. Instead, we make use of a simpler procedure
to determine an estimate for the CO-to-H$_{2}$ conversion factor, denoted as 
$X_{\rm CO, est}$. We first select a set of independent sightlines through our simulation, one per
resolution element. We next compute H$_{2}$ and CO column densities along each of these 
sightlines. We convert each of the CO column densities into an estimate of the optical depth of the
gas in the CO $J = 1 \rightarrow 0$ transition, under the assumptions that (a) the CO level populations 
are in LTE, (b) the gas is isothermal, with a temperature equal to the CO-weighted mean temperature found in the actual simulation, and (c) the CO linewidth is uniform, and is given by $\Delta v = 3 \: {\rm km} \: {\rm s^{-1}}$. Given an estimate for the CO optical depth, we then compute an estimate for
$W_{\rm CO}$ using the same technique as in Pineda {\em et al.\/}~(\cite{pineda08}). Finally, we average over all the sightlines to compute a mean intensity $\meanW$ for the simulation, and do 
the same for the H$_{2}$ to arrive at a mean H$_{2}$ column density. $X_{\rm CO, est}$ is then
simply the ratio of these two values.

\begin{figure}
\centering
\includegraphics[width=14pc, angle=270]{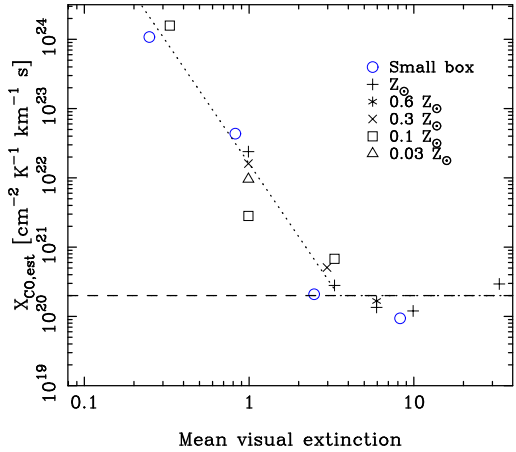}   
\caption{Estimate of the CO-to-H$_{2}$ conversion factor $X_{\rm CO, est}$,
plotted as a function of the mean visual extinction of the gas, $\meanAV$. 
At $\meanAV > 3$, the values we find are consistent with the value of 
$X_{\rm CO} = 2 \times 10^{20} {\rm cm^{-2} \: K^{-1} \: km^{-1} \: s}$
determined observationally for the Milky Way by Dame {\em et~al.\/} (\cite{dame01}), 
indicated in the plot by the horizontal dashed line. 
At $\meanAV < 3$, we find evidence for a strong dependence of
$X_{\rm CO, est}$ on $\meanAV$. The empirical fit given by 
Equation~\ref{emp-fit} is indicated as the dotted line in the Figure, and demonstrates that
at low $\meanAV$, the CO-to-H$_{2}$ conversion factor increases roughly
as $X_{\rm CO, est} \propto A_{\rm V}^{-3.5}$. 
\label{xfact}
}
\end{figure}

Using this procedure, we have computed $X_{\rm CO, est}$ for each of our simulations.
Figure~\ref{xfact} shows how the values we obtain depend on the mean visual extinction
of the gas, $\meanAV$. We see from the figure that there is a clear change in the behaviour 
of $X_{\rm CO, est}$ at $\meanAV \sim 3$. In clouds with larger mean extinctions, 
$X_{\rm CO, est}$ is roughly constant and has a value consistent with the 
observationally-determined value for the Milky Way. 
On the other hand, for smaller mean extinctions, $X_{\rm CO, est}$ increases
sharply with decreasing $\meanAV$. This behaviour is caused by a rapid fall-off in the CO
abundance with decreasing mean extinction, which leads to a corresponding sharp drop
in $W_{\rm CO}$. Because it does not self-shield efficiently, CO molecules are protected
from photodissociation primarily by dust extinction, and so as this decreases, the CO 
abundance decreases much more rapidly than the H$_{2}$ abundance, resulting in the
rapid increase we find for $X_{\rm CO, est}$. The dependence of $X_{\rm CO, est}$ on
$\meanAV$ can be described by the empirical fitting function
\begin{equation}
X_{\rm CO, est} \simeq \left \{
\begin{array}{lr}
2.0 \times 10^{20} & A_{\rm V} > 3.5 \\
2.0 \times 10^{20} \left(A_{\rm V} / 3.5\right)^{-3.5} 
& A_{\rm V} < 3.5
\end{array}
\right .
\label{emp-fit}
\end{equation}
illustrated in Figure~\ref{xfact} by the dotted line.

The sharp fall-off in $W_{\rm CO}$ with decreasing mean extinction has an important
consequence. In order to detect CO emission from GMCs in  Local Group galaxies, the integrated
intensity of the emission must be  $\sim 1 \, {\rm K} \, {\rm km} \, {\rm s^{-1}}$ or higher, and we
find in our simulations that only the clouds with $\meanAV > 1$ have integrated intensities above
this value. Therefore, GMCs detectable in CO will always sit on the right-hand side of Figure~\ref{xfact},
in the regime where $X_{\rm CO, est}$ is roughly constant, providing a simple explanation for why
values of $X_{\rm CO}$ in extragalactic systems derived using CO observations are always roughly 
the same as the Galactic value. On the other hand, GMCs with mass determinations that do not rely on CO are not constrained to fall in the regime where $W_{\rm CO}$ is large, and so may occur 
anywhere in the plot, thereby explaining why they are often found to have X-factors that are much 
larger than the Galactic value. A further consequence of the behaviour of $X_{\rm CO}$ and 
$W_{\rm CO}$ is the prediction that CO-bright clouds in low metallicity systems must be larger 
and/or denser than their Milky Way counterparts, since at lower metallicity a larger surface density
of gas is required to produce the necessary mean extinction.


\end{document}